\documentclass{article}
\usepackage{fortschritte}
\usepackage{epsfig}
\usepackage{amssymb}
\def\beq{\begin{equation}}                     %
\def\eeq{\end{equation}}                       %
\def\bea{\begin{eqnarray}}                     
\def\eea{\end{eqnarray}}                       
\newcommand{\mbb}{\mathbb}
\begin {document}                 
\begin{flushright} \vspace{-2cm} 
{\small MPP-2004-155 \\ 
hep-th/0412025} \end{flushright}
\def\titleline{
%
%
%
Recent Progress in  Intersecting D-brane Models
%
}
\def\email_speaker{
{\tt 
%
%
speaker@engine.institute.country             
}}
\def\authors{
%
%
%
%
%
Ralph  Blumenhagen 
}
\def\addresses{
%
%
%
%
Max-Planck Institut f\"ur Physik, \\
F\"ohringer Ring 6, 80805 M\"unchen, Germany                           %
}
\def\abstracttext{
%
%
The aim of this article is to review some recent progress in the 
field of intersecting D-brane models.  This  includes  the construction
of chiral, semi-realistic flux compactifications,
the systematic study of Gepner model orientifolds,
the computation of various terms in the low energy effective action
and the investigation of the statistics
of solutions to the tadpole cancellation conditions.

}
 \large
\makefront

\section{Introduction}

Since its discovery in the year 2000 \cite{IBW}, the field of intersecting D-brane
models has developed and matured considerably and is by now
a well established string theoretic framework for (semi-)realistic
string compactifications (for reviews see \cite{review}). 
While in the beginning most of the effort went into developing
new model building techniques and studying simple examples, during the last
2 years  one of the main questions  was the computation of 
the low energy effective action and its physical consequences.
At least for flat string backgrounds now techniques
are available to compute the tree-level K\"ahler potential 
\cite{Lust:2004cx}, the Yukawa couplings 
\cite{yukawas,Cremades:2003qj,Cremades:2004wa}, 
gauge threshold corrections \cite{Lust:thres} and
the susy breaking soft terms  on D3 and D7-branes 
\cite{Camara:2003ku,Lust:2004fi,Kane:2004hm}.

Independently  we have learned that compactifications with
non-vanishing background fluxes give rise to a 
scalar potential, which allows to freeze some of the
moduli notoriously present in supersymmetric
string compactifications \cite{Giddings:2001yu}. 
These observations not
only led to a proposal for realizing metastable de-Sitter vacua
in string theory \cite{Kachru:2003aw} 
but also have drastically influenced 
the way we think about the landscape of string vacua.
In fact, the number of vacua appearing in 
such flux compactifications is so enormous that
without any further guidance
it is quite questionable whether there is any chance 
to ever find {\it the} realistic string vacuum.
Alternatively, a statistical approach to the string
vacuum problem was proposed in \cite{Douglas:2003um}
and methodologically  pioneered in \cite{Ashok:2003gk}.

In this article I would like to briefly summarize  some of the
recent developments in the field of intersecting
D-brane models. Please note  that the selection of
topics reflects the author's preferences and is 
not meant to disgrace other important contributions  in this field. 
In section 2, I will review  recent attempts to combine
flux compactifications with intersecting branes by
constructing semi-realistic 
chiral flux compactifications in a treatable setting.
Section 3 is devoted to summarize recent studies on
a large set of intersecting D-brane models on 
highly curved backgrounds, namely orientifolds of
Gepner models. 
In section 4, I will try to briefly sketch   what has been achieved 
in determining the low energy effective action for intersecting
D-brane models. 
Finally, section 5 surveys a very recent investigation
of the statistics of solutions to the tadpole
cancellation conditions.
This article is based on a review talk the author gave
at the 37$^{\rm th}$ Symposium Ahrenshoop, 23-27 August 2004,
and as such not only
contains results published by the author himself but also
by many others.

\section{Semi-realistic flux compactifications}

During the last years string compactifications with non-trivial
background fluxes were studied in many variations.
It turned out that these models can solve some of the
problems purely geometric string compactification notoriously
had. 
In particular, certain fluxes
induce an effective potential that still possesses
supersymmetric minima, which allows to freeze (some of)
the moduli generically appearing in string theory. 
It was also possible 
to break supersymmetry in a controlled way by turning on additional
internal flux components. Finally,
by taking also some non-perturbative
effects into account, for the first time
strong evidence was given  that non-supersymmetric meta-stable
de-Sitter vacua do exist in string theory \cite{Kachru:2003aw}.

However, the original framework contained only parallel D3-branes
on which no semi-realistic gauge theory can arise. 
To reconcile this,  it was proposed to combine
ideas from flux compactifications with ideas from intersecting
respectively magnetised branes to build chiral semi-realistic
string models with fluxes and partly frozen moduli 
\cite{Cascales:2003zp}.
Concretely, the Type IIB closed string background was chosen
to be the orbifold $M=T^6/\mbb{Z}_2\times \mbb{Z}_2$ with Hodge numbers
$(h_{21},h_{11})=(51,3)$ (the T-dual of the Type IIA
orientifold studied in \cite{csu}). In addition one performs
the orientifold projection $\Omega R(-1)^{F_L}$, where
$R$ reflects all six internal directions. 
Flux compactifications on the mirror symmetric Calabi-Yau
given by the orbifold with discrete torsion were also
considered and will be discussed in more detail in \cite{bcms}.

Turning on 3-form fluxes $H_3$ and $F_3$, the Chern-Simons
term in the Type IIB effective action induces a 
4-form tadpole given by
\beq
    N_{flux}={1\over (4\pi^2\alpha')^2 }
\int_M H_3\wedge F_3.
\eeq
The fluxes obey the Bianchi identity and take values in $H^3(M,\mbb{Z})$, i.e.
\beq
    {1\over (2\pi)^2 \alpha' }  \int_M H_3\in N_{min} \mbb{Z}, \quad\quad
    {1\over (2\pi)^2 \alpha' }  \int_M F_3\in N_{min} \mbb{Z},
\eeq
where $N_{min}$ is an integer guaranteeing  that in orbifold
models only untwisted 3-form fluxes are turned on, for which we
can trust the supergravity approximation. Taking also
the orientifold projection into account for  
the $T^6/\mbb{Z}_2\times \mbb{Z}_2$
orbifold one gets $N_{min}=8$.
Defining $G_3=\tau H_3 + F_3$ the kinetic term for the G-flux
\beq
   S_G=-{1\over 4\kappa_{10}^2 \Im(\tau)} \int_M   G\wedge \star_6
  G,
\eeq
generates a scalar potential which can be derived from the
GVW-superpotential \cite{Gukov,Taylor:1999ii}
\beq
\label{GVW}
W=\int_M \Omega_3\wedge G.
\eeq
As is apparent the scalar potential only depends on the complex
structure moduli and the dilaton. 
It is of no-scale type and vanishes for imaginary self-dual
fluxes (ISD) $\star_6 G=i\, G$, e.g. for $G$  of type 
$(2,1)$ or $(0,3)$. One can show that the minimum is supersymmetric
if $G$ is solely of type $(2,1)$. 

In order to cancel the resulting tadpoles, one introduces
in the usual way magnetised D9-branes, which are T-dual
to the intersecting D6-branes studied in many papers.
Such a magnetised brane is characterised by three pairs
of integers $(n^I_a,m^I_a)$ which satisfy
\beq
    {m^I_a\over 2\pi}\int_{T^2_I}   F_a^I = n^I_a, 
\eeq
where the $m^I_a$ denote the wrapping number
of the D9-brane around the torus $T^2_I$ and
$n^I_a$ is the magnetic flux. 
The orientifold projection acts as follows on these
quantum numbers $\Omega R(-1)^{F_L}:(n^I_a,m^I_a)\to (n^I_a,-m^I_a)$.
Since $h_{11}=3$ one gets in the orientifold four tadpole
cancellation conditions
\bea
\label{tad}
\sum_a N_a\, n^1_a\, n^2_a\, n^3_a &=& 8-{N_{flux}\over 4} \\
\sum_a N_a\, n^I_a\, m^J_a\, m^K_a &=& -8\quad {\rm for}\ I\ne J\ne K\ne I .
\eea
In order for each brane to preserve the same supersymmetry as the
orientifold planes, they have to satisfy
\beq
\sum_I \arctan\left(  {m^I_a {\cal K}^I\over n^I_a} \right) =0,
\eeq
where ${\cal K}^I$ denotes the volume of the I-th torus $T^2$
in units of $\alpha'$.
The number of chiral fermions between two different magnetised
branes is given by the index
\beq
    I_{ab}=\prod_I (n_a^I\, m_b^I -m_a^I\, n_b^I)  
\eeq
and can lead to matter in bifundamental, symmetric or anti-symmetric
representations of the gauge group.

Taking the flux quantisation with $N_{min}=8$ into account, the
contribution of the flux to the D3-brane tadpole is given by
$N_{flux}/4\in 16\, \mbb{Z}$. Therefore, for non-trivial flux the
right hand side of the D3-brane tadpole cancellation condition (\ref{tad})
is always negative. In \cite{Cascales:2003zp}
 this led to the conclusion that
no globally supersymmetric solutions to the tadpole cancellation
conditions  do exist. However this was too naive, namely in \cite{Marchesano:2004yq}
it was shown that there exist supersymmetric branes which 
give the "wrong" sign in one of the four tadpole cancellation
conditions. 
Consider for instance the magnetised brane
$(n_a^I,m_a^I)=(-2,1)(-3,1)(-4,1)$, which is supersymmetric for
\beq
\arctan (A_1/2)+\arctan (A_2/3)+\arctan (A_3/4)=\pi
\eeq
and contributes as $(-24,-4,-2,-3)$ to the four tadpole
conditions. 
Precisely branes of this type were used in \cite{Marchesano:2004yq,Cvetic:2004xx}
to construct   globally supersymmetric, chiral, MSSM like flux compactifications.
For illustrative purposes, let me present here only one 
of their examples.

Choosing the 3-form flux as
\beq
G_3={8\over \sqrt 3}\, e^{-{\pi i\over 3}}\, ( d \overline{z} _1 dz_2 dz_3 +
        dz_1 d \overline{z} _2 dz_3 +   dz_1 dz_2 d \overline{z} _3 ),
\eeq 
yields a contribution $N_{flux}/4= 48$ to the tadpole
condition and freezes the moduli at $U^I=\tau=e^{2\pi i/3}$.
Introducing the supersymmetric branes shown in Table 1 cancels
all the tadpoles and gives rise to a one-generation
MSSM-like model with gauge group
\beq
   G=SU(3)\times SU(2)\times SU(2)\times U(1)_{B-L}\times
    [ U(1)'\times USp(8) ].
\eeq
Supersymmetry enforces the additional constraint 
$A_2=A_3$. 
 For more technical and phenomenological details of such models
please consult  the original literature.
\begin{table}
\caption{Wrapping numbers for semi-realistic model.} 
\centering
\vspace{3mm}
\label{twrap}
\begin{tabular}{|c||c|c|c|}
\hline
$N_a$ & $(n^1_a,m^1_a)$ & $(n^2_a,m^2_a)$ & $(n^3_a,m^3_a)$  \\
\hline\hline
$N_a=3$ &  $(1,0)$ & $(1,1)$ & $(1,-1)$ \\
\hline
$N_b=1$ &  $(0,1)$ & $(1,0)$ & $(0,-1)$ \\
\hline
$N_c=1$ &  $(0,1)$ & $(0,-1)$ & $(1,0)$ \\
\hline
$N_d=1$ &  $(1,0)$ & $(1,1)$ & $(1,-1)$ \\
\hline
\hline
$N_{h_1}=1$ &  $(-2,1)$ & $(-3,1)$ & $(-4,1)$ \\
\hline
$N_{h_2}=1$ &  $(-2,1)$ & $(-4,1)$ & $(-3,1)$ \\
\hline
$N_f=4$ &  $(1,0)$ & $(1,0)$ & $(1,0)$ \\
\hline
\end{tabular}
\end{table}
Note that the branes $b,c$ can be placed directly on top
the corresponding $O7$-planes yielding a gauge group
$SU(2)\times SU(2)$.
This example shows that it is indeed possible to construct
supersymmetric semi-realistic string models with fluxes
and partly frozen moduli. This is an encouraging
observation, but of course much more work is needed
to really establish an entire class of such models. 
A first step towards generalizations of these kinds
of models has been carried out in \cite{Font:2004cy}.
A different approach allowing  more general gauge fluxes
on $T^6$ has been proposed in \cite{anto}.

\section{Gepner Model orientifolds}

After some earlier attempts \cite{Angelantonj:1996mw},
 there was quite a revival of interest
in the construction of Gepner model orientifolds during the last
year \cite{Aldazabal:2003ub,Blumenhagen:2003su,Brunner:2004zd,Dijkstra:2004ym}. 
Since Gepner  models are known to describe exactly solvable 
non-linear sigma models on certain Calabi-Yau manifolds
at radii of  string scale size, the study of such models tells
us something about a regime  which is not accessible 
via perturbative supergravity methods. I would like to
emphasize  that while the construction of supersymmetric
intersecting brane models in the geometric regime
of the Calabi-Yau moduli space is hampered by
our ignorance about concrete examples of special Lagrangian 3-cycles,
at small radii conformal field theory techniques applied
to Gepner models  allow us to construct a plethora 
of concrete intersecting D-brane models. 

One starts with Type IIB string theory on a Calabi-Yau threefold, where
the internal part is described by a GSO projected  ${\cal N}=2$ superconformal
field theory (SCFT) with central charge $c=9$. Gepner proposed to 
use tensor products of unitary representations of the ${\cal N}=2$ 
super Virasoro algebra for the internal SCFT, which has
to be equipped with a GSO projection onto states of integer
$U(1)$ charge. There are 168 tensor products of this sort,
which can be identified with certain Calabi-Yau spaces
given by hypersurfaces in weighted projective spaces.
However, by additional orbifold respectively simple current
extensions, this set of modular invariant SCFTs can be extended
to $O(1000)$ different models.

Next one defines an orientifold projection $\Omega \sigma$, where
$\Omega$ denotes the word-sheet parity transformation
and $\sigma$ an internal symmetry of the SCFT, like phase
symmetries or extra quantum symmetries acting differently
on various twisted sectors. 
There have been different approaches to proceed with the
orientifold construction. 

First, following the earlier
attempts in \cite{Angelantonj:1996mw}, 
one can implement the orientifold
projection and compute directly the loop-channel
Klein-bottle amplitude \cite{Aldazabal:2003ub,Blumenhagen:2003su}. 
This allows one to fix
the crosscap state up to some sign factors, which
have to be determined later.
From the crosscap states one can read off the various
non-vanishing tadpoles, which are to be cancelled 
by additional sources given by rational
boundary states (D-branes) of the Gepner model. 
Note that these highly symmetric rational boundary
states \` a la Cardy are only a small subset of all 
possible supersymmetric boundary states. It would
be interesting to construct more general boundary states
and use them for model building.
Requiring consistency between the boundary and
crosscap states allows one to fix the
signs in the crosscap states. It only remains
to solve  the tadpole cancellation conditions,
which for a complicated model can become
quite involved by hand. 

In the second approach one invokes known results from
the study of the crosscap states in general
rational conformal field theory \cite{Fuchs:2000cm,Brunner:2003zm}. 
Therefore one starts
with the crosscap  and boundary states and from there determines
the rest of the model \cite{Brunner:2004zd,Dijkstra:2004ym}. 
Clearly both approaches are  equivalent and historically
both have been used.

The main advantage of this SCFT approach is that everything
is algebraic and the resulting expressions for the
tadpole conditions, the annulus and M\"obius strip amplitudes
can be written in  general form admitting the 
implementation in a computer code. 
In fact in  \cite{Dijkstra:2004ym} a systematic computer search for three-generation
semi-realistic models was carried out. This impressive search is by far
the most ambitious attempt carried out ever to search   for Standard-like
intersecting D-brane models. It revealed that there exist
$O(10^5)$ of models resembling  the Standard Model at least
in their  topological quantities. 
The authors also investigated  other phenomenological features 
like the number of Higgs fields, the number of adjoint scalars,
the tree-level gauge couplings and the number of hidden sector gauge groups. 
For details I would like to refer the readers to the original work 
\cite{Dijkstra:2004ym}.

It would be interesting to see how many of these Standard-like
models survive, if  one also requires  more refined data
to agree with the Standard Model. As an example one should develop
techniques to compute Yukawa couplings, i.e. the three point
function of three boundary changing operators. 
Since all radii are fixed at the order of the string scale, it is far from
clear how a hierarchy between the Yukawa's could be generated. 
It could be like proposed in \cite{Cremades:2003qj}
 that only the third generation is massive
at string tree level,  and that the lighter two generations
get their mass via loop corrections.

In view of the statistical approach to the string vacuum problem, to be
discussed in section 5, the set of Gepner model orientifolds 
can clearly  provide a nice testing ground. 
It would be interesting to develop statistical methods to
determine the various distributions of gauge theoretic observables
in the ensemble of Gepner model orientifolds and test the
results against a Monte-Carlo based frequency computation.

A question which appears to be much harder to answer is what happens
when one turns on fluxes in the Gepner model. 
Since Gepner models live at string size radii, the supergravity techniques one usually
invokes  to study flux compactifications are not applicable. 
Moreover, one cannot simply take the Gepner models and continuously 
deform them to large radii 
for they are lying on  lines of marginal stability and supersymmetry is generally broken
by the deformation. 
Therefore, before even thinking about fluxes, one has to study what happens
for deformations away from the Gepner point.

\section{The low energy effective action}

In this section I would like to briefly summarize the status
of knowledge about the low energy effective action arising
from supersymmetric intersecting D-brane models. 
Only a very tiny selection of the relevant material is given and
I would like to refer the reader for more details to the 
still growing original literature. 

It is clear that in order to make contact between stringy constructions
and low energy particle physics, 
the determination  of the low energy effective action is absolutely
essential. Since the precise form  of the action depends sensitively 
on the details of the string model, it is first important
to work out  the necessary general technical tools and 
apply them to sufficiently simple toy models. 
This describes precisely the situation at the moment, where
we are learning how such computations are performed for
relatively simple  examples.

\subsection{The supersymmetric effective action}

The effective ${\cal N}=1$ supersymmetric action depends on
the holomorphic gauge kinetic function
 $f(Z)$, the holomorphic superpotential $W(Z)$
and the non-holomorphic K\"ahler potential $K(Z,\o Z)$.

\begin{itemize}
\item{Gauge couplings: Each stack of D6-branes comes with its own
     gauge coupling. For a supersymmetric brane wrapping a 3-cycle
     $\pi_a$, the tree level result for the holomorphic gauge kinetic function
     reads \cite{bbkl,Blumenhagen:2003jy}
\beq
   f(U_i)={M_s^3\over (2\pi)^4}\left[ e^{-\varphi} \int_{\pi_a}
     \Re (\Omega_3) +2i \int_{\pi_a} C_3 \right]
\eeq
where $C_3$ denotes the R-R three-form. Apparently, the gauge coupling
at tree level only depends on the complex structure moduli.
It is known that the gauge couplings receive corrections
at the one-loop level, the so-called gauge threshold correction.
These are very hard to compute for a generic model, as one has to know
the massive string spectrum. For intersecting branes on 
a toroidal (orbifold) background these
threshold corrections have been computed in \cite{Lust:thres}.
The main result is that the one-loop correction for open string
sectors preserving ${\cal N}=4$ supersymmetry vanish. For ${\cal N}=2$
sectors they both depend on the complex and K\"ahler moduli, 
whereas for ${\cal N}=1$ they are given by the nice expression
for the gauge coupling of $SU(N_a)$
\beq
  \Delta_{ab}=-b_{ab} \ln{\Gamma (1-\theta_{ba}^1)
        \Gamma(1-\theta_{ba}^2)
          \Gamma(1+\theta_{ba}^1+\theta_{ba}^2)\over
         \Gamma(1+\theta_{ba}^1)
        \Gamma(1+\theta_{ba}^2)
          \Gamma(1-\theta_{ba}^1-\theta_{ba}^2)},
\eeq
where $b_{ab}=N_b I_{ab} {\rm Tr}(Q^2_a)$ and
$\theta^I_{ba}={1\over \pi}\Phi_{ba}^I$ with
$\Phi_{ba}^I$ denoting  the difference of the intersection angles
on the three $T^2$ factors. Implicitly, this expression only depends
on the complex structure moduli.
}
\item{K\"ahler potential: The K\"ahler potential for the various 
massless chiral multiplets arising in intersecting D-brane models
has been determined in \cite{Lust:2004cx} by a string scattering computation.
Again for concreteness it was assumed that one is dealing with  a toroidal 
(orbifold)  background.
Here I would like to present only one of the many nice results
contained in \cite{Lust:2004cx}, namely the K\"ahler potential for the charged
chiral superfields. These fields arise from open strings stretched
between two intersecting D-branes and the K\"ahler potential
reads
\beq
     G_{ab}=\kappa_4^{-2} \prod_{I=1}^3 (T^I - \overline{T} ^I)^{-\theta^I_{ab}} 
         \, \sqrt{ {\Gamma( \theta_{ab}^I)\over 
              \Gamma(1- \theta_{ab}^I)}}        
\eeq
depending on the K\"ahler moduli and implicitly also on the
complex structure moduli.
}
\item{The superpotential: In section 2, we have already encountered
one possible contribution to the superpotential. 
Namely considering the T-dual Type IIB set-up with magnetised
D-branes, turning on 3-form flux $G_3$ induces the  GVW-type
superpotential (\ref{GVW}). 
Another contribution contains the Yukawa couplings for three
massless chiral multiplets. The resulting physical Yukawa
couplings, i.e. those for canonical normalized fields 
are given by \cite{yukawas}
\bea
      Y_{ijk}&=&e^{K/2}\, (K_{ab} K_{bc} K_{ca} )^{-{1\over 2}}\, W_{ijk}\\
          &=&2\pi \prod_{I=1}^3 \left[ 16\pi^2 
       {\Gamma(1-\theta^I)\Gamma(1-\nu^I)\Gamma(\theta^I+\nu^I)\over
        \Gamma(\theta^I)\Gamma(\nu^I)\Gamma(1-\theta^I-\nu^I) } 
        \right]^{1\over 4}\, \sum_m \exp\left( -{A_I(m)\over 2\pi\alpha'}
       \right)
\eea
with $\theta^I={1\over \pi}\Phi^I_{ba}$ and $\nu^I={1\over \pi}\Phi^I_{cb}$. 
The sum is over all world-sheet instantons with the boundary given
by the three intersecting D-branes. These sums have been
analysed in \cite{Cremades:2003qj}, where it was shown that they can be expressed
in terms of theta-functions. It was shown in \cite{Cremades:2004wa} 
that in the mirror symmetric
model with magnetised branes the superpotential part $W_{ijk}$ 
of the Yukawa couplings can be obtained by a tree-level (in $\alpha'$)
computation.
}
\end{itemize}

\subsection{Soft susy breaking terms}

For phenomenological reasons it is clearly not sufficient to determine
the supersymmetric effective action. Since supersymmetry must be broken
at a certain scale, it is desirable to also have control
over a supersymmetry breaking mechanism and compute the resulting
soft supersymmetry breaking terms. 
For  supersymmetry breaking via fluxes in Type IIB string theory
with magnetised D-branes such a program has been carried out recently
\cite{Camara:2003ku,Lust:2004fi,Kane:2004hm}.
Here two different approaches have been pursued: One can either
include the effect of fluxes in the Dirac-Born-Infeld action
for the D3 and D7 branes and expand the resulting action
to lowest order in the transverse coordinates  and read off the resulting
soft terms. Alternatively, one can parameterize the susy breaking
by the VEVs of the various auxiliary fields of the closed string
moduli superfields and use the general supergravity formalism for 
determining the resulting
soft-terms. These two approaches are expected to be equivalent. 
Again for more detailed results the reader should consult the
original literature. Here I would like to just state some
general observations:
\begin{itemize}
\item{The soft terms vanish for $D3$-branes in ISD flux backgrounds.
      However, for $\o{D3}$ branes the soft terms are non-vanishing
     and in particular, masses are induced for the brane position
     moduli.}

\item{The susy breaking soft terms on $D7$ branes vanish for $(2,1)$ fluxes,
     but are non-vanishing for a susy breaking $(0,3)$ component
    of the flux. For $(2,1)$ fluxes  also supersymmetric 
$\mu$-terms for the brane moduli are
     generated, which is in accord with the F-theory expectation \cite{lars}.  }

\item{The induced susy breaking scale on the D7-branes is roughly
    of the order $M_s^2\over M_{pl}$ and therefore 
    for TeV scale susy breaking favours a string scale in the intermediate
     regime.}
\end{itemize}

\section{Statistics of intersecting D-branes }

After the realization that flux compactifications give rise to 
 a densely populated 
landscape of string vacua, M.R. Douglas was the first to propose 
that, given these $10^{500}$ different vacua, one should better try
a statistical approach to the string vacuum problem \cite{Douglas:2003um}. 
It was pointed out that such an approach  might even turn out to be predictive in the sense
that it leads for instance to strong statistical correlations having
 the potential to  falsify string theory.

In \cite{Ashok:2003gk,Denef:2004cf} very powerful statistical methods were developed
to determine to distribution of flux vacua over the complex
structure moduli space. These were extended and refined
in \cite{DeWolfe:2004ns} to also count the number of Minkowskian backgrounds. 
Of course for phenomenological reasons it is  very
important to also include the brane sector into the statistics. 
Some general claims were  made already in the original work
\cite{Douglas:2003um}, 
while more refined methods were presented in  \cite{Blumenhagen:2004xx}
in the context of intersecting D-branes respectively magnetized D-branes.

More concretely in \cite{Blumenhagen:2004xx}, for the ensemble of intersecting branes
on certain toroidal orientifolds, the statistical distribution
of various gauge theoretic quantities was studied, like the rank of the gauge group, 
the number of models with an $SU(M)$ gauge factor and the number
of generations. In \cite{Blumenhagen:2004xx} the examples 
of intersecting branes on the $T^2$, $T^4/\mbb{Z}_2$ and  
$T^6/\mbb{Z}_2\times \mbb{Z}_2$
orientifolds were discussed.

Let us briefly summarize the main computational
technique used to determine these distributions.
The first step is to determine all or at least a large,
preferably  representative subset of supersymmetric branes.
After solving the supersymmetry constraints, in all the  examples
discussed in \cite{Blumenhagen:2004xx}, 
this was given by a subset $S$ of the naively allowed wrapping numbers $X_I$.
As a constraint one faces the various tadpole cancellation
conditions
\begin{equation}
       \sum_{a=1}^k  N_a\,  X_{a,I} = L_I 
\end{equation}
with $I=1,\ldots,b_3/2$  and $L_I$ denoting the contribution
from the orientifold planes and the fluxes.
Realizing that the problem of counting the number of solutions
to these equations is similar in spirit to the counting of
unordered partitions of an integer, the method used
for determining the approximate expansion in the latter case
can be generalized to our problem. 
It turns out that the number of such solutions  is given by the expression
\bea
\label{numb}
  {\cal N}(\vec L ) 
                     &\simeq& {1\over (2\pi i)^{b_3\over 2}} 
    \oint \prod_I {dq_I \over q_I^{L_I+1} }\, 
                \exp\left(\sum_{X_I\in S} {\prod_I q_I^{X_I}
                \over 1-\prod_I q_I^{X_I}} \right),                     
\eea
which can be  evaluated at leading order by a saddle point
approximation with 
\begin{equation}
       f(\vec q)= \sum_{X_I\in S} {\prod_I q_I^{X_I}
                \over 1-\prod_I q_I^{X_I}} - \sum_I (L_I+1)\, \log q_I. 
\end{equation}
The saddle point is determined by the condition 
$\nabla f(\vec q)|_{\vec q_0}=0$,
and the second order saddle point approximation reads
\begin{equation}
{\cal N}^{(2)}(\vec L)=  {1\over  \sqrt{2\pi}^n}\, { e^{f(\vec q_{0})} 
         \over 
    \sqrt{ \det\left[ \left( {\partial^2 f\over \partial q_i\partial q_j}\right) 
        \right]_{q_0}}}.  
\end{equation}
For the more complicated statistical distributions, one gets similar
results, which all can be estimated via the saddle-point approximation.

After having checked that the number of solutions to the tadpole
cancellation conditions for the 8D, 6D and 4D examples are finite,
various gauge theoretic distributions were computed and compared to
a brute force computer classification. 
Cutting a long story short, the following qualitative results
we obtained
\begin{itemize}
\item{The probability to find an $SU(M)$ gauge factor scales
like 
\bea
P(M)\simeq \exp\left( -\sqrt{\log L\over L} M \right)
\eea
and for $\sum_{i=1}^k M_i\ll L$ satisfies mutual independence, i.e.
$P(M_1\, \ldots M_k)=\prod_i P(M_i)$.
}
\item{The rank distribution yields approximately a Gauss curve
with the maximum depending on the complex structure moduli and
whether one allows multiple wrapping or not.
} 
\item{Defining a measure for the chirality of a solution by
    $\chi=\pi'\circ \pi$,  in the 6D case a scaling 
like 
\bea
P(\chi)\simeq \exp\left( -\kappa\,  \sqrt{\chi}  \right)
\eea
was found 
with $\kappa$ denoting some constant depending presumably on the $L_I$. 
}
\item{A  strong statistical correlation between the rank
of the gauge group and the chirality (number of families) was found,
which
can be traced back to the tadpole cancellation conditions.
The higher the chirality is, the lower becomes
the average rank of the gauge group. 
}
\end{itemize}

It is interesting to study what happens
when one combines the flux statistics with the
D-brane statistics. The question is whether by averaging over
more parameters maybe the various distributions become
essentially uniform. To get a first glimpse,
the T-dual Type IIB model with fluxes and magnetized
D9-branes was considered. 
For the rank distribution for instance one  gets the following
expression
\begin{equation}
  \overline{P}(r)={1\over N_{norm}} 
\sum_{N_{flux}=0}^{N_{flux}^{max}}
          {(N_{flux}+1)^K}\,\, {\cal N}(r;L_0-N_{flux},L_1,L_2,L_3), 
\end{equation}
where ${\cal N}(r,L_I)$ is just the unnormalised part
of the distribution ({\ref{numb}).
Varying the number of 3-cycles, $K$,  gives the distribution shown in
Figure 1.      
\begin{figure}
\begin{center}
\epsfbox{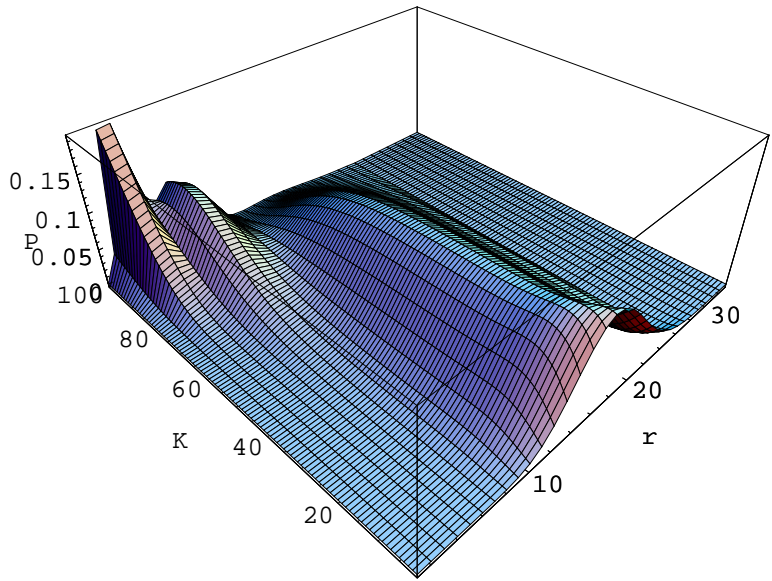}
\caption{The rank distribution after averaging over flux vacua for $L_0=L_1=L_2=L_3=8$,
$U_I=1$ and $N_{flux}^{max}=11$.}
\label{fct2}
\end{center}
\end{figure}
One realizes that the distribution is far from being uniform and
that for large values of 3-cycles new maxima appear, which
for instance contain models of the sort discussed in section 2. 

Clearly, we are just beginning to approach the problem of unravelling
the statistics on the landscape of string theory. The final
aim would be to perform the statistics over as many parameters
as possible to really get a realistic picture of what overall statistical
averages can tell us about the distribution of various
physical quantities.
The methods shown above might play an important role whenever one
encounters string theoretic constraints similar
to the tadpole cancellation conditions.
More modestly, as a next step it would be interesting to
study the distributions of heterotic string vacua and to see
whether, as expected from string dualities,
they feature  similar patterns as the orientifolds.
As I have mentioned already, Gepner model orientifolds
might provide a nice testing ground for comparing 
and possibly refining the technical statistical tools. 
In principle, having agreed upon a good statistical ensemble one would
like to address questions concerned directly with the
Standard Model, like 
\begin{itemize}
\item{What is the percentage of models having the right gauge group,
       matter and number of families?}
\item{How drastically is this number reduced by requiring more
     detailed constraints, like the right gauge and Yukawa couplings,
     the right Higgs couplings, absence of exotic matter?}
\item{Having installed all phenomenological constraints, how
     does the distribution of the susy breaking scale and
    the cosmological constant look like?}
\end{itemize}     
The answers to these questions will  strongly depend on possible
statistical correlations among the various quantities, 
the realization of which
I consider as  the most interesting aspect of this endeavour.

\vskip 0.2cm
{\bf Acknowledgement} I would like to thank
F. Gmeiner, G. Honecker, D. L\"ust, S. Stieberger, 
T. Taylor  and T. Weigand for collaboration
on part of the work presented in this article.
I would also like to thank M. Cvetic, F. Marchesano and
G. Shiu for useful discussions. 
Moreover, I would like to thank the organizers of the
37$^{\rm th}$ Symposium Ahrenshoop, 23-27 August 2004  for asking me
to give an overview talk on intersecting D-brane models. 



\begin{thebibliography}{77}


\bibitem{IBW}
R.~Blumenhagen, L.~G\"orlich, B.~K\"ors and D.~L\"ust,
JHEP {\bf 0010}, 006 (2000)
[arXiv:hep-th/0007024];
C.~Angelantonj, I.~Antoniadis, E.~Dudas, A.~Sagnotti, 
Phys. Lett. B {\bf 489}, 223 (2000)
[arXiv:hep-th/0007090];
G.~Aldazabal, S.~Franco, L.~E.~Ibanez, R.~Rabadan, A.~M.~Uranga,
JHEP {\bf 0102}, 047 (2001) 
[arXiv:hep-ph/0011132];
G.~Aldazabal, S.~Franco, L.~E.~Ibanez, R.~Rabadan, A.~M.~Uranga,
J.\ Math.\ Phys.\  {\bf 42}, 3103 (2001)
[arXiv:hep-th/0011073].

\bibitem{review}
A.~M.~Uranga,
Class.\ Quant.\ Grav.\  {\bf 20}, S373 (2003)
[arXiv:hep-th/0301032];
D.~L{\" u}st,
Class.\ Quant.\ Grav.\  {\bf 21}, S1399 (2004)
[arXiv:hep-th/0401156];
E.~Kiritsis,
Fortsch.\ Phys.\  {\bf 52}, 200 (2004)
[arXiv:hep-th/0310001].

\bibitem{Lust:2004cx}
D.~L{\" u}st, P.~Mayr, R.~Richter and S.~Stieberger,
Nucl.\ Phys.\ B {\bf 696}, 205 (2004)
[arXiv:hep-th/0404134].


\bibitem{Cremades:2003qj}
D.~Cremades, L.~E.~Ibanez and F.~Marchesano,
JHEP {\bf 0307}, 038 (2003)
[arXiv:hep-th/0302105].

\bibitem{Cremades:2004wa}
D.~Cremades, L.~E.~Ibanez and F.~Marchesano,
JHEP {\bf 0405}, 079 (2004)
[arXiv:hep-th/0404229].


\bibitem{Lust:thres}
D.~L{\" u}st and S.~Stieberger,
arXiv:hep-th/0302221.


\bibitem{yukawas}
M.~Cvetic and I.~Papadimitriou,
Phys.\ Rev.\ D {\bf 68}, 046001 (2003)
[Erratum-ibid.\ D {\bf 70}, 029903 (2004)]
[arXiv:hep-th/0303083];
S.~A.~Abel and A.~W.~Owen,
Nucl.\ Phys.\ B {\bf 663}, 197 (2003)
[arXiv:hep-th/0303124].

\bibitem{Camara:2003ku}
P.~G.~Camara, L.~E.~Ibanez and A.~M.~Uranga,
Nucl.\ Phys.\ B {\bf 689}, 195 (2004)
[arXiv:hep-th/0311241];
P.~G.~Camara, L.~E.~Ibanez and A.~M.~Uranga,
arXiv:hep-th/0408036;
L.~E.~Ibanez,
arXiv:hep-ph/0408064.


\bibitem{Lust:2004fi}
D.~L{\" u}st, S.~Reffert and S.~Stieberger,
arXiv:hep-th/0406092;
D.~L{\" u}st, S.~Reffert and S.~Stieberger,
arXiv:hep-th/0410074.

\bibitem{Kane:2004hm}
M.~Grana, T.~W.~Grimm, H.~Jockers and J.~Louis,
Nucl.\ Phys.\ B {\bf 690}, 21 (2004)
[arXiv:hep-th/0312232];
G.~L.~Kane, P.~Kumar, J.~D.~Lykken and T.~T.~Wang,
arXiv:hep-ph/0411125.




\bibitem{Giddings:2001yu}
K.~Dasgupta, G.~Rajesh and S.~Sethi,
JHEP {\bf 9908} (1999) 023
[arXiv:hep-th/9908088];
S.~Gukov,
Nucl.\ Phys.\ B {\bf 574}, 169 (2000)
[arXiv:hep-th/9911011];
S.~B.~Giddings, S.~Kachru and J.~Polchinski,
Phys.\ Rev.\ D {\bf 66}, 106006 (2002)
[arXiv:hep-th/0105097];
S.~Kachru, M.~B.~Schulz and S.~Trivedi,
JHEP {\bf 0310}, 007 (2003)
[arXiv:hep-th/0201028].

\bibitem{Kachru:2003aw}
S.~Kachru, R.~Kallosh, A.~Linde and S.~P.~Trivedi,
Phys.\ Rev.\ D {\bf 68}, 046005 (2003)
[arXiv:hep-th/0301240].


\bibitem{Douglas:2003um}
M.~R.~Douglas,
JHEP {\bf 0305}, 046 (2003)
[arXiv:hep-th/0303194].

\bibitem{Ashok:2003gk}
S.~Ashok and M.~R.~Douglas,
JHEP {\bf 0401}, 060 (2004)
[arXiv:hep-th/0307049];
M.~R.~Douglas,
arXiv:hep-ph/0401004;
F.~Denef and M.~R.~Douglas,
JHEP {\bf 0405}, 072 (2004)
[arXiv:hep-th/0404116];
M.~R.~Douglas,
arXiv:hep-th/0409207.



\bibitem{Cascales:2003zp}
R.~Blumenhagen, D.~L{\" u}st and T.~R.~Taylor,
Nucl.\ Phys.\ B {\bf 663}, 319 (2003)
[arXiv:hep-th/0303016];
J.~F.~G.~Cascales and A.~M.~Uranga,
JHEP {\bf 0305}, 011 (2003)
[arXiv:hep-th/0303024];
J.~F.~G.~Cascales and A.~M.~Uranga,
arXiv:hep-th/0311250.

\bibitem{csu}
M.~Cvetic, G.~Shiu and  A.~M.~Uranga,  
Nucl. Phys. B {\bf 615}, 3  (2001) 
[arXiv:hep-th/0107166].



\bibitem{bcms}
R.~Blumenhagen, M.~Cvetic, F.~Marchesano and G.~Shiu,
{\it in preparation}.



\bibitem{Gukov}
S.~Gukov, C.~Vafa and E.~Witten,
Nucl.\ Phys.\ B {\bf 584}, 69 (2000)
[Erratum-ibid.\ B {\bf 608}, 477 (2001)]
[arXiv:hep-th/9906070].



\bibitem{Taylor:1999ii}
T.~R.~Taylor and C.~Vafa,
Phys.\ Lett.\ B {\bf 474}, 130 (2000)
[arXiv:hep-th/9912152].

\bibitem{Marchesano:2004yq}
F.~Marchesano and G.~Shiu,
arXiv:hep-th/0408059;
F.~Marchesano and G.~Shiu,
arXiv:hep-th/0409132;
F.~Marchesano, G.~Shiu and L.~T.~Wang,
arXiv:hep-th/0411080.


\bibitem{Cvetic:2004xx}
M.~Cvetic and T.~Liu,
arXiv:hep-th/0409032.

\bibitem{Font:2004cy}
A.~Font,
arXiv:hep-th/0410206.


\bibitem{anto}
I.~Antoniadis and T.~Maillard,
arXiv:hep-th/0412008.


\bibitem{Angelantonj:1996mw}
C.~Angelantonj, M.~Bianchi, G.~Pradisi, A.~Sagnotti and Y.~S.~Stanev,
Phys.\ Lett.\ B {\bf 387}, 743 (1996)
[arXiv:hep-th/9607229];
R.~Blumenhagen and A.~Wisskirchen,
Phys.\ Lett.\ B {\bf 438}, 52 (1998)
[arXiv:hep-th/9806131].



\bibitem{Aldazabal:2003ub}
G.~Aldazabal, E.~C.~Andres, M.~Leston and C.~Nunez,
JHEP {\bf 0309}, 067 (2003)
[arXiv:hep-th/0307183];
G.~Aldazabal, E.~C.~Andres and J.~E.~Juknevich,
JHEP {\bf 0405}, 054 (2004)
[arXiv:hep-th/0403262].

\bibitem{Blumenhagen:2003su}
R.~Blumenhagen,
JHEP {\bf 0311}, 055 (2003)
[arXiv:hep-th/0310244];
R.~Blumenhagen and T.~Weigand,
JHEP {\bf 0402}, 041 (2004)
[arXiv:hep-th/0401148];
R.~Blumenhagen and T.~Weigand,
Phys.\ Lett.\ B {\bf 591}, 161 (2004)
[arXiv:hep-th/0403299].

\bibitem{Brunner:2004zd}
I.~Brunner, K.~Hori, K.~Hosomichi and J.~Walcher,
arXiv:hep-th/0401137.


\bibitem{Dijkstra:2004ym}
T.~P.~T.~Dijkstra, L.~R.~Huiszoon and A.~N.~Schellekens,
arXiv:hep-th/0403196;
T.~P.~T.~Dijkstra, L.~R.~Huiszoon and A.~N.~Schellekens,
arXiv:hep-th/0411129.

\bibitem{Fuchs:2000cm}
J.~Fuchs, L.~R.~Huiszoon, A.~N.~Schellekens, C.~Schweigert and J.~Walcher,
Phys.\ Lett.\ B {\bf 495}, 427 (2000)
[arXiv:hep-th/0007174].


\bibitem{Brunner:2003zm}
I.~Brunner and K.~Hori,
arXiv:hep-th/0303135.



\bibitem{bbkl}
D.~Cremades, L.~E.~Ibanez and F.~Marchesano,
JHEP {\bf 0207}, 009 (2002)
[arXiv:hep-th/0201205];
R.~Blumenhagen, V.~Braun, B.~K\"ors and D.~L\"ust,
JHEP {\bf 0207}, 026 (2002) 
[arXiv:hep-th/0206038].



\bibitem{Blumenhagen:2003jy}
R.~Blumenhagen, D.~L{\" u}st and S.~Stieberger,
JHEP {\bf 0307}, 036 (2003)
[arXiv:hep-th/0305146].

\bibitem{lars}
L.~ G\"orlich, S.~Kachru, P.~K.~Tripathy and S.~P.~Trivedi,
arXiv:hep-th/0407130.




\bibitem{Denef:2004cf}
F.~Denef and M.~R.~Douglas,
arXiv:hep-th/0411183.

\bibitem{DeWolfe:2004ns}
O.~DeWolfe, A.~Giryavets, S.~Kachru and W.~Taylor,
arXiv:hep-th/0411061.

\bibitem{Blumenhagen:2004xx}
R.~Blumenhagen, F.~Gmeiner, G.~Honecker, D.~L{\" u}st and T.~Weigand,
arXiv:hep-th/0411173.


\bibitem{Giryavets:2004zr}
A.~Giryavets, S.~Kachru and P.~K.~Tripathy,
JHEP {\bf 0408} (2004) 002
[arXiv:hep-th/0404243];
J.~P.~Conlon and F.~Quevedo,
JHEP {\bf 0410} (2004) 039
[arXiv:hep-th/0409215].



\end{thebibliography}
\end{document}